\documentclass[twoside,letterpaper]{article}
\usepackage{amsmath}
\usepackage{amssymb}
\usepackage{cite}
\usepackage{graphicx}

\input ijmpblat
\newcommand{\nn}{\nonumber}

\newcommand{\kb}{k_{_{\mathrm{B}}}}
\newcommand{\eps}{\varepsilon}
\newcommand{\bp}{\mathbf{p}}
\newcommand{\la}{\left<}
\newcommand{\ra}{\right>}

\newcommand{\Tc}{\ensuremath{T_{\mathrm{c}}}}
\newcommand{\D}{\displaystyle}
\newcommand{\jour}[4]{{\nineit #1}\ {\ninebf #2},\ #3\ (#4)}
\newcommand{\Fref}[1]{Fig.~\ref{#1}}
\newcommand{\Eref}[1]{Eq.~(\ref{#1})}
\newcommand{\Rref}[1]{Ref.~\citen{#1}}
\newcommand{\PRL}{Phys. Rev. Lett.}
\newcommand{\PR}{Phys. Rev.}

\begin{document}
\runninghead{T. Mishonov \& E. Penev}%
            {Effective masses of disordered BCS superconductors}
%
\thispagestyle{empty}
\setcounter{page}{1}
%
%
\vspace*{0.88truein}
\fpage{1}
\centerline{\textbf{{BERNOULLI POTENTIAL, HALL CONSTANT AND COOPER PAIRS }}}
\vspace*{0.035truein}
\centerline{\textbf{EFFECTIVE MASSES IN DISORDERED BCS SUPERCONDUCTORS}}
\vspace*{0.37truein}
\centerline{\footnotesize TODOR
MISHONOV$^{\dag,}$\footnote{Permanent address$^\ddag$,
   Corresponding author; phone: (++32) 16 327183, fax: (++32) 16 327983,\\
   e-mail:~\texttt{todor.mishonov@fys.kuleuven.ac.be}
               }\ \
  {\normalsize and} EVGENI PENEV$^{\ddag}$
}
\vspace*{0.015truein}
\centerline{$^{\dag}$\footnotesize\it
   Laboratorium voor Vaste-Stoffysica en Magnetisme, %
   Katholieke Universiteit Leuven
}
\baselineskip=10pt
\centerline{\footnotesize\it
   Celestijnenlaan 200 D, B-3001 Leuven, Belgium
}
\vspace*{0.015truein}
\centerline{$^\ddag$\footnotesize\it
   Faculty of Physics, Sofia University ``St. Kliment Ohridski''
}
\baselineskip=10pt
\centerline{\footnotesize\it
    5~J.~Bourchier Blvd., 1164 Sofia, Bulgaria
}
\vspace*{0.225truein}

\vspace*{0.21truein}
\abstracts{It is analyzed what fundamental new information for the
properties of the superconductors can be obtained by systematic
investigation of the Bernoulli effect. It is shown that it is a
tool to determine the effective mass of Cooper pairs, the volume
density of charge carriers, the temperature dependence of the
penetration depth and condensation energy. The theoretical results
for disordered and anisotropic gap superconductors are
systematized for this aim. For clean-anisotropic-gap
superconductors is presented a simple derivation for the
temperature dependence of the penetration depth}{}{}

\vspace*{0.21truein}

\vspace*{1pt}\textlineskip
\section{Introduction}
\label{sec:1}
\vspace*{-0.5pt}
\noindent

The Landau theory of second-order phase transitions\cite{LLV} and its
realization for superconductors, the Ginzburg-Landau (GL) gauge
theory,\cite{GL} can be classified as belonging to the most
illuminating theoretical achievements in XXth-century physics. The
basic concepts\cite{LL:9} advanced in these theories often find applications in
interdisciplinary research. The microscopic Bardeen-Cooper-Schrieffer
(BCS) theory\cite{bcs:57} makes it possible to calculate the
parameters of the GL theory. Thus the phenomenology of
superconductivity can be reliably derived once the parameters of the
microscopic Hamiltonian are specified. Such a scheme ensures that
there is no missing link between the microscopic theory and the
material properties of the superconductors. On the other hand
Bernoulli effect is one of the oldest effects known in the physics
which in as a result of the energy conservation can be realized
strictly speaking only for superfluids. For superconductor the
Bernoulli theorem gives a relation between current density and the
current induced contact potential difference at constant in thermal
equilibrium electrochemical potential.

The purpose of the present paper is to systematize the results for
the GL theory of disordered and anisotropic gap superconductors
and to point out what new information for the fundamental
parameters of superconductors can be extracted using the Bernoulli
effect. We are describing in short how the experiments can be done
and how the effective mass of Cooper pairs, the volume density of
charge carriers and London penetration depth can be extracted from
investigation of the Bernoulli effect. The proposed experiments
require only standard low frequency electronic measurements. In
parallel we analyze a simple methodical  derivation of the London
penetration depth for clean superconductors. We will start our
analysis with the GL equation, then we will analyze disorder
renormalization and experimental accessibility of the
effective Cooper pair mass, we will analyse the Bernoulli effect
and temperature dependence of the penetration depth. Finally we
will conclude which experimental development we consider as most
perspective.

\section{Ginzburg-Landau equation}
\label{sec:2}
\vspace*{-0.5pt}
\noindent

Consider now the gradient terms of the Ginzburg-Landau (GL) theory.  A
general expression for the tensor of the squared coherence lengths
$(\hat{\xi}^2)_{\alpha\beta}$ has been derived
by Pokrovsky and Pokrovsky\cite{Pokrovsky:96}:
\begin{align}
\label{eq:CoherenceLength}
(\hat{\xi^2})_{\alpha\beta} = & \frac{\hbar^2}{\left(4\pi\kb
    \Tc\right)^2}\frac{1}{\la\chi^2\ra}
    \Bigl( \zeta_{3,0}\la v_\alpha v_\beta\chi^2\ra\\
    & + 2x_c\zeta_{3,1}\la v_\alpha
    v_\beta\chi\ra \la\chi\ra  + x_c^2\zeta_{3,2}\la v_\alpha
    v_\beta\ra\la\chi\ra^2 \Bigr),\nn
\end{align}
where
\begin{equation}
\zeta_{k,l} \equiv \zeta_{k,l}(x_c+1/2),\quad
\zeta_{k,l}(z)=\sum_{n=0}^{\infty}(n+z)^{-k}(n+1/2)^{-l},
\quad x_c=\frac{\hbar}{2\pi k_\mathrm{B}\tau(T_c)},
\label{eq:zkl}
\end{equation}
$\tau(T_c)$ is the electron scattering time,
$ v_\alpha$ are the components of the velocity, and
$\langle\dots\rangle$ means averaging on the Fermi surface.
In terms of $\hat{\xi^2}$ the GL equation for the space-dependent
order parameter $\Xi(\mathbf{r})$ can be written as
\begin{equation}
\Bigl(- \mathbf{D}\cdot\hat{\xi^2}\cdot\mathbf{D}
 +\epsilon + \frac{b}{a_0}\left|\Xi\right|^2 \Bigr) \Xi=0,
\label{eq:GL}
\end{equation}
where
\begin{alignat}{2}
 \mathbf{D} &= \frac{\partial}{\partial\mathbf{r}}
 -\frac{2\pi}{\Phi_0} \mathbf{A}(\mathbf{r}), \qquad
  & \Phi_0 &= 2\pi\hbar/e^*,\\
 |e^*|  & = 2|e|, \qquad & \epsilon  &= \frac{T-\Tc}{\Tc}.
\end{alignat}
Equation~(\ref{eq:GL}) is the extremum (minimum) condition for the GL
free energy
\begin{equation}
 F[\Xi,\mathbf{A}] = \int d\mathbf{r} \biggl\{-\frac{\hbar^2}{2}
  [\mathbf{D}\Xi(\mathbf{r})]^* \cdot \tilde M^{-1}
   \cdot\mathbf{D}\Xi(\mathbf{r})  +a(T) |\Xi(\mathbf{r})|^2 +
   \frac{1}{2}b |\Xi(\mathbf{r})|^4\biggr\},
\label{eq:GLfreeEnergy}
\end{equation}
where
\begin{equation}
(\tilde M^{-1})_{\alpha\beta} =
\frac{2a_0}{\hbar^2} (\hat{\xi^2})_{\alpha\beta}, \text{ and } a(T)=\epsilon a_0.
\label{eq:tildeMass}
\end{equation}
Using the eigenvalues $\xi_\alpha$ and $\tilde M_\alpha $ of the
tensors $\hat{\xi^2}$ and $\tilde M$ we can introduce the
temperature-dependent GL coherence lengths and penetration depths,
respectively ($0< -\epsilon \ll 1$):
\begin{equation}
\xi_{\alpha,\mathrm{GL}}(T) \approx\xi_{\alpha}/\sqrt{-\epsilon},\quad
\lambda_{\alpha,\mathrm{GL}}(T) \approx\lambda_\alpha/\sqrt{-\epsilon},
\end{equation}
which satisfy the following GL relations
\begin{equation}
\frac{1}{\lambda_\alpha^2} =\frac{(a_0/b)e^{*2}}{c^2\varepsilon_0}\frac{1}{\tilde M_\alpha}
 = \frac{2 a_0^2 e^{*2}}{\hbar^2 \varepsilon_0 c^2 b} \xi_\alpha^2
 = 8\pi^2\mu_0\frac{a_0^2\xi_\alpha^2}{b\Phi_0^2}
 = \frac{8\pi^2\mu_0}{\Phi_0^2}\,T_c\Delta C\xi_\alpha^2,
\end{equation}
where in Gaussian units $\mu_0=4\pi$ and $\Delta C=a_0^2/T_c b$ is the jump of
the specific heat at $T_c$ per unit volume. The explicit formulae
for the GL coefficients $a_0$ and $b$ are given in Ref.\cite{Pokrovsky:96}

\section{Isotropic alloys. Disorder renormalization of the Cooper pair mass}

Let us illustrate now the operation of the above general expressions
on the important for the applications case of dirty isotropic
alloys. In this case we have
\begin{alignat}{3}
\la v_\alpha v_\beta\ra & = \frac{1}{3} v_\mathrm{F}^2\delta_{\alpha\beta},
 \quad &\chi_{\bp} &=\mathrm{const},\nn\\
 (\hat{\xi^2})_{\alpha\beta} &= \xi^2 \delta_{\alpha\beta}
 \quad & \tilde M_{\alpha\beta} &= \tilde M\delta_{\alpha\beta}.
\end{alignat}
Using the identity
\begin{equation}
\zeta_{3,0}+2x_c\zeta_{3,1}+x_c^2\zeta_{3,2}=\zeta_{1,2}
\end{equation}
one obtains the classical result due to Gor'kov,
see the textbook \Rref{Ketterson:99}
\begin{equation}
\xi^2=\frac{1}{3}\left(\frac{\hbar
v_\mathrm{F}}{4\pi\kb\Tc}\right)^2\zeta_{1,2}
=\xi_\mathrm{clean}^2\frac{\tilde M_\mathrm{clean}}{\tilde M(x)}.
\end{equation}
Here we wish to recall some notation used in
different works on the physics of superconuctivity
(see, e.g., \Rref{Tinkham:96}):
\begin{align}
\xi_\mathrm{clean} & = \frac{\xi_0}{4}\frac{2\pi}{\gamma}\sqrt{\frac{7\zeta(3)}{12}}
  =\frac{\xi_0}{4}\frac{2\Delta(0)/\kb\Tc}{\sqrt{\Delta C/C_n(\Tc)}}
  = 0.738\xi_0\approx\xi_{0\mathrm{P}},\nn\\ 
\xi_0 & = \frac{\hbar v_\mathrm{F}}{\pi\Delta(0)}
    = \frac{\gamma}{\pi^2}\frac{\hbar v_\mathrm{F}}{\kb\Tc}
    = 0.1805\frac{\hbar v_\mathrm{F}}{\kb\Tc},\nn\\ 
2\Delta(0)&=\frac{2\pi}{\gamma}\kb\Tc =3.53\kb\Tc,
    \quad \gamma= 1.781\dots,
    \nn\\ 
    \frac{\xi^2}{\xi_\mathrm{clean}^2}&
    =\frac{\lambda_\mathrm{clean}^2}{\lambda^2}
    =\frac{\tilde M_\mathrm{clean}}{\tilde M(x)}
    =\frac{\zeta_{1,2}\left(x_c+\frac{1}{2}\right)}{7\zeta(3)}
    \approx\frac{1}{1+\frac{14\zeta(3)}{\pi^2}x_c}
    =\frac{1}{1+\frac{\xi_{0\mathrm{P}}}{l}},\label{eq:Pippard}\\
\xi_{0\mathrm{P}}&
    =\frac{7\zeta(3)}{2\pi\gamma}\xi_0=0.752\xi_0 
    =\frac{7\zeta(3)}{2\pi^3}\frac{\hbar v_\mathrm{F}}{\kb\Tc}\nn\\
    &=0.1357\frac{\hbar v_\mathrm{F}}{\kb\Tc} 
    =\frac{2\sqrt{21\zeta(3)}}{\pi^2}\xi_\mathrm{clean}=1.018\xi_\mathrm{clean},\nn 
\end{align}
and finally $l=v_\mathrm{F}\frac{\tau}{2}.$
%
\begin{figure}[t]
\centering
\includegraphics[width=0.6\columnwidth]{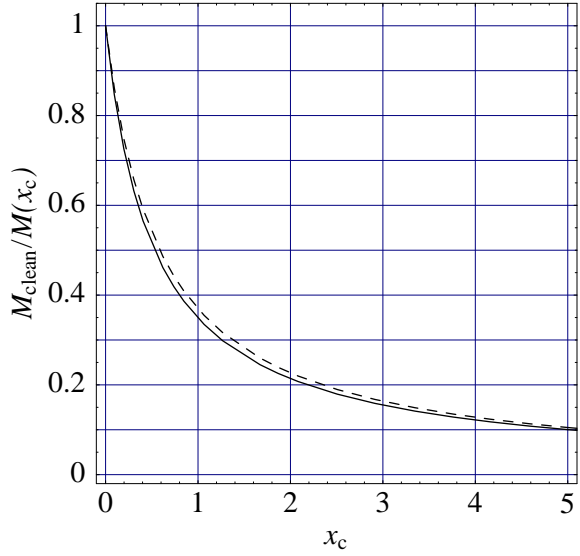}\\
\caption{\small Disorder renormalization of the Cooper pair mass
$\tilde M/\tilde M_\mathrm{clean} =\xi_{\mathrm{clean}}^2/\xi^2
=\lambda^2/\lambda^2_{\mathrm{clean}} \approx
H_{c2}/H_{c2,\mathrm{clean}}$ vs. dimensionless scattering rate
$x_c=\hbar/2\pi\kb\Tc\tau(\Tc)$. The exact microscopic result
\protect\Eref{eq:Pippard} is shown by the full line. The dashed line
is the Pad\'e approximant derived\protect\cite{Mishonov:94a} within
framework of the Pippard-Landau theory.
\label{fig:1}}
\end{figure}
%
The 2\% difference between the length $\xi,$ introduced by Ginzburg
and Landau,\cite{GL,LL:9} and $\xi_{0\mathrm{P}}$ introduced by
Pippard,\cite{Pippard:53} is experimentally inaccessible. So is the
several percent difference between the exact microscopic result
(\ref{eq:Pippard}) and its Pad\'e approximant shown in
\Fref{fig:1}. The Pad\'e approximant for the mass renormalization can
be easily derived\cite{Mishonov:94a} within framework of the
Pippard-Landau theory, i.e.  using the generalization of the local GL
theory and nonlocal Pippard electrodynamics. These are
generalized\cite{Mishonov:94a} by inserting the Pippard kernel between
the two gradients in the GL expression for the free energy
(\ref{eq:GLfreeEnergy}):
\begin{gather}
\iint d\mathrm{r}\, d\mathrm{r}' \sum_{\alpha,\beta}
 [D_{\alpha}\Xi(\mathbf{r})]^* \delta(\mathbf{R}) D_\beta\Xi(\mathbf{r}')\longrightarrow\\
\iint d\mathrm{r}\, d\mathrm{r}' [\mathbf{D}\Xi(\mathbf{r})]^* \cdot
 \frac{ 3 \mathbf{R}\otimes \mathbf{R}}{4\pi\xi_{0\mathrm{P}}R^4}\,
 \mathrm{e}^{-R/\xi_{\mathrm{P}}} \cdot \mathbf{D}  \Xi(\mathbf{r}^\prime),\nn
\end{gather}
where $\mathbf{R}=\mathbf{r}-\mathbf{r}',$ $\xi\approx\xi_{\mathrm{P}},$ and
\begin{equation}
\frac{1}{\xi_{\mathrm{P}}}=\frac{1}{\xi_{0\mathrm{P}}}+\frac{1}{l},\quad
\xi_\mathrm{GL}(T)\approx\frac{\xi_{\mathrm{P}}}{\sqrt{|\epsilon|}}.
\end{equation}

\section{Experimental accessibility of the effective Cooper pair
mass $(T=0)$}
\label{sec:BE}

\begin{figure}[t]
\centering
\includegraphics[width=0.8\columnwidth]{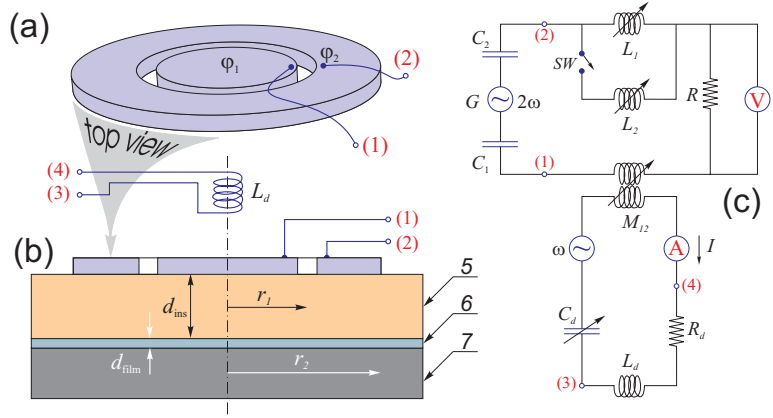}\\
\caption{\small Cooper pair mass spectroscopy based on the Bernoulli
potential (after \Rref{Mishonov:94b}). (a) Top view (b) Cross section,
(c) Equivalent electric scheme. Two electrodes, circle- (1) and
ring-shaped electrode (2), should be produced on the insulating layer
capping the superconducting film. (3) and (4) denote the contacts of
the drive coil with inductance $L_d$ and resistance $R_d$;
(5)---insulator layer of thickness $d_{\mathrm{ins}};$
(6)---superconducting film with thickness $d_{\mathrm{film}} <
\lambda_{ab}(0)$; (7)---substrate; $M_{12}$---mutual inductance;
$l_1,$ $l_2$---variable inductances; $r$---load resistor;
\textcircled{V}---voltmeter; \textcircled{A}---ammeter;
$SW$---switch; $C_d$---capacitor of the drive resonance contour with
resonance frequency $\omega;$ $G$---Bernoulli voltage generator with
doubled frequency 2$\omega;$ $C_1,\; C_2$---capacitances between the
superconducting film and the metal electrodes (1) and (2).
\label{fig:2}
}
\end{figure}

The tensor introduced in \Eref{eq:tildeMass} acquires dimension of
mass if we renormalize the GL order parameter so as to have dimension
of density. The total charge carrier density $n_{\mathrm{tot}}$ is a
fundamental theoretical notion. Thus it is of principle interests
whether this quantity is experimentally accessible through the
equilibrium thermodynamic properties of the vortex-free
Meissner-Ochsenfeld phase. An approach to this problem has been
outlined in \Rref{Mishonov:94b} and is based on the current-induced
contact-potential difference at the surface of a
superconductor\cite{Mishonov:94b} (we call it the London-Hall effect):
\begin{equation}
\label{eq:bulk}
\Delta \varphi = -{\cal R}_\mathrm{H}\frac{B^2}{2\mu_0}.
\end{equation}
For superconductors characterized with local electrodynamics
$\Delta\varphi$ is a manifestation of the energy conservation law and
the Bernoulli theorem\cite{Mishonov:94b} for charged superfluids in
thermodynamic equilibrium:
\begin{equation}
\Delta \varphi=-\frac{1}{2\varepsilon_0c^2}{\cal R}_\mathrm{H}\lambda^2(T)j^2.
\label{eq:BTH}
\end{equation}
For $T=0$ this equation reads as one fluid energy conservation low
\begin{equation}
\frac{1}{2}M_\mathrm{CP}v^2+e^*\Delta\varphi=0,
\end{equation}
where $M_\mathrm{CP}$ is the effective mass of Cooper pairs.
For anisotropic superconductors $\lambda$ in \Eref{eq:BTH} corresponds
to the direction of the current density $\mathbf{j}$ at the
superconductor surface.  All radio frequencies are too small compared
to the typical gap parameters of superconductors.  Hence
$\Delta\varphi$ can be measured using a lock-in amplifier in
conjunction with a suitable low-noise preamplifier.  The
superconductor surface under investigation, as shown in \Fref{fig:2},
plays the role of one of the plates of the capacitor with capacitance
$c$ and resistance $r_c$.

The condition under which $\Delta\varphi$ can be directly measured
reads
\begin{equation}
C f_{\text{lock-in}} \gg \frac{1}{R_C} + \frac{1}{R_{\text{lock-in}}},
\end{equation}
where $f_{\text{lock-in}}$ and $R_{\text{lock-in}}$ are, respectively,
the operating frequency and the internal resistance of the lock-in
amplifier. Typical voltages are $\sim$nV but for high-\Tc\
superconductors the Bernoulli signals will be considerably
stronger. The key technological problem lies in the the properties of
the superconductor-insulator interface---its quality should be
comparable to that of the samples used for investigation of electric
effects in superconductors.

\begin{figure}[t]
\centering
\includegraphics[width=0.8\columnwidth]{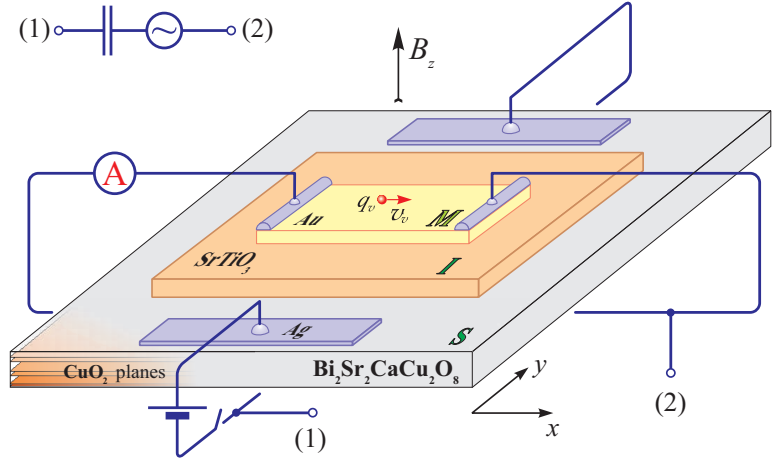}\\
  \caption{\label{fig:setup}
  \textit{Gedanken} set-up proposed in \Rref{Mishonov:00}
  to determine the vortex charge
  and the Cooper pair mass. Thin Bi$_2$Sr$_2$CaCu$_2$O$_8$ layer is
  thread by perpendicular magnetic field $B_z$. The voltage $V_y$ applied through
  the Ag electrodes in circuit (1) creates a drift of the vortices
  with mean velocity $v_v.$ Due to the Bernoulli effect the superfluid
  currents around every vortex create a change in the electric
  potential on the superconducting surface. The Bernoulli potential
  of the vortex leads to an electric polarization on the normal Au
  surface. The charge $q_v$, related to the vortex, has the same drift
  velocity $v_v.$ The corresponding current $I_x$ in
  circuit (2) can be read by a sensitive ammeter. The quality of
  the SrTiO$_3$ plate should be high enough so as to allow detection of
  the interface Hall current without being significant perturbed by
  the leakage currents between circuits (1) and (2).
}
   \end{figure}

For pure crystals of elemental metals the total charge density is
accessible trough Hall-effect measurements in the normal phase, but
for dirty superconductors only the London-Hall effect gives such a
possibility. Determination of the Hall constant by the current-induced
contact-potential difference provides a tool for determination of the
superfluid density at zero temperature $n(0),$ $en_{\mathrm{tot}} =
e^*n(0)=1/{\cal R}_{\mathrm{H}}.$ Thus knowing $n(0)$ and
$\lambda(0)$ one can determine the effective mass of the Cooper pairs
$M_\mathrm{CP}$:
\begin{equation}
\frac{1}{\lambda^2(0)}=\frac{n(0)e^{*2}}{M_\mathrm{CP}c^2\varepsilon_0},
\qquad M_\mathrm{CP}=\frac{e^*\lambda^2(0)}{c^2\varepsilon_0{\cal R}_\mathrm{H}}.
\end{equation}
All the three important parameters, ${\cal R}_{\mathrm{H}}$,
$M_\mathrm{CP},$ and $\lambda(T),$ can be determined by measuring the
contact potential difference in thin ($d_{\mathrm{film}}\ll
\lambda(T)$) and thick ($d_{\mathrm{film}}\gg \lambda(T)$)
superconducting films.  We should mention that strong enhancement of
the effective mass due to disorder is valid for arbitrary band
anisotropy if the averaged gap does not vanish:
\begin{equation}
(\hat \xi^2_{\mathrm{dirty}})_{\alpha\beta} \propto \frac{\tau}{\Tc}
    \frac{\la\chi\ra^2}{\la\chi^2\ra}
    \la v_\alpha v_\beta\ra, \qquad (x \gg 1).
\end{equation}
As it was commented earlier,\cite{Mishonov:91} the effective mass
of the Cooper pairs was first measured by Fiory \textit{et
al.}\cite{Fiory:91} using electrostatic doping of
YBa$_2$Cu$_3$O$_{7-\delta}$ films. This pioneering experiment
confirmed that the effective mass shows weak temperature
dependence. Certainly, there are other methods\cite{Mishonov:00}
for absolute determination of $M_\mathrm{CP},$ e.g., from the
surface Hall effect.\cite{Mishonov:99} For the layered cuprates
the problem of determining the Cooper pair mass is related to
problem of the vortex charge.\cite{Mishonov:00} The aim of the
model experiment suggested in \Rref{Mishonov:00} was to check
whether the Bernoulli potential of the pancake vortices creates an
electrical polarization necessary to explain the Hall effect in
the vortex creep regime. The supposed experimental setup is
sketched in Fig.~\ref{fig:setup}. Up to now the is only one
experimental hint concerning the electrostatic polarization around
the vortices, the NMR and NQR study by Kumagai, Nozaki and
Matsuda.\cite{Kumagai:01}
%

Finally we wish to mention that electric polarization of a
metal-oxide-superconductor plane capacitor by a sudden light
impulse can give a the condensation energy
\begin{equation}
\label{Eq:CPD T}
\frac{\Delta \varphi}{{\cal R}_\mathrm{H}} = \frac{B_c^2(T)}{2\mu_0}.
\end{equation}
We have to take into account that electric polarization is related
to work function, workfunction is related to the chemical
potential which is the Gibbs free energy per particle and in such
a way we have an electric method for measurement of the
thermodynamic critical magnetic field $B_c(T).$

\section{What has to be done. A short working program for the experiment}

After the analysis of the theory let us describe what has to be
done experimentally for creation of Cooper pair mass spectroscopy
based on the Bernoulli effect. Among the other electric field
effect in superconductors such as surface Hall effect and
electrostatic modulation of the kinetic inductance the Bernoulli
effect looks simplest and experimental research is better to start
with him. The main advantage is that in this case the electric
field is very small and there are no restrictions related to the
brake-trough voltage of the insulating layer. It is necessary
insulator to be good only for small voltages. These investigations
can be a by-product of preliminary research of the superconductor
insulator interfaces in further trial to make superconducting
field effect transistors. The first step will be qualitatively
observation of the Bernoulli effect in already existing
superconductor structures. A numerical calculation of the current
distribution in the framework of London electrodynamics can
convert this observation in quantitative measurement. The simplest
possible set-up is a superconducting nanobridge in which an
appropriate gate can be performed. In this field effect transistor
type structure\cite{Mishonov:94b} current in the strip can be
excited by capacitor connections id source and drain area. In such
a way generation of the current harmonics by the contacts can be
avoided.

Later on a systematic investigation can be started. The main
purpose is to try to use already prepared big number of
superconductor films without any destroying patterning to be
carried out. The gold Bernoulli electrodes should be evaporated on
a thin SrTiO$_3$ substrate. The investigated film should be
affixed on this plate. A small coil, or a solenoid with diameter
of order of 1 mm, for example, will excite the eddy currents. The
detector circuit should be switched between the circular and ring
electrodes patterned on the insulator plate, cf. Fig.~\ref{fig:2}.
The cross talk between detector and drive coil should be annulated
by a variable mutual inductance. The harmonic current should be
applied to the drive coil and the Bernoulli signal will be
detected at second harmonics using a low noise preamplifier before
the loch-in. A resonance technique additionally can use the high-Q
factor to suppres the noise and signal at basic frequency.

In parallel a numerical simulation of the experimental setup will
allow the result of measurements to be delivered as material
constants. After a systematic investigation of the collected films
would be possible to extract fundamental information of the doping
dependence of the effective mass of Cooper pairs in cuprate
CuO$_2$ superconductors and the almost temperature independent
London-Hall coefficient.

Finally the experimental setup could be elaborated as a commercial
device (superfluid densitometer) intended for fast contactless
investigation of the Bernoulli effect in superconductors. Such a
device can find application for the monitoring of the quality of
the thin superconducting films used in second-generation
superconducting cables, for example. The investigation of other
electric field effects in the superconductors will be the next
step of the investigation of electric field effect in
superconductors.

The measurement of the Bernoulli signal in irradiated thin films
can become a routine tool for investigation of these disordered
superconductors.

Except for the Cooper pair mass spectroscopy and charge carriers
densitometry Bernoulli effect, i.e. measurement of the contact
potential difference can become unique tool for determination of
the condensation energy especially for type-II superconductors.
For sudden heating we can use laser light illumination of the
set-up for the Bernoulli measurements.

\section{Temperature dependence of the penetration depth and optical mass}

Let us attempt now a simple derivation of the temperature
dependence of the penetration depth for clean superconductors. Our
objective is to show that the result is in agreement with the
formula for the coherence length, \Eref{eq:CoherenceLength}.

For a superconductor at $T=0$, consider the current response to a
small space-homogeneous variation of the vector potential
$\mathbf{A}(t).$ The Fermi surface is shifted in the momentum
space as a rigid object; see, e.g., \Rref{Kogan:02}. In this
respect there is no principal difference between a superconductor
and a normal metal at optical frequencies. The difference is only
that the optical approach, $\omega\tau\gg1,$ can be used for the
static response of the clean superconductor. The electromagnetic
field brings about a small change of the electron momentum
$\mathbf{Q}=-e\mathbf{A}(t)$ and the shift in the momentum space,
$\bp \rightarrow \bp + \mathbf{Q},$ creates in turn a small shift
of the electron kinetic energies:
\begin{gather}
\eps_\bp \rightarrow\varepsilon_{\bp + \mathbf{Q}}\approx
    \varepsilon_\bp + \mathbf{v}_\bp \cdot \mathbf{Q}+
    \frac{1}{2}\mathbf{Q}\cdot m^{-1}_\bp \cdot \mathbf{Q},\\
 m^{-1}_{\bp}= \frac{\partial \mathbf{v}_{\bp}}{\partial\bp}
         = \frac{\partial^2\eps_\bp}{\partial\bp^2}.
\end{gather}
Thus for the increase of the kinetic energy density we obtain:

\begin{align}
\label{eq:OpticalResponse}
 w_{\text{kin}} &= \frac{1}{2}\mathbf{Q}\cdot
    m^{-1}\cdot\mathbf{Q}\,n_c
    =\frac{1}{2}\mathbf{v}_\mathrm{dr}\cdot
    m\cdot\mathbf{v}_\mathrm{dr}n_c,\\ & =
    \frac{1}{2}\varepsilon_0 c^2 \sum_{\alpha,\beta=1}^{3}
    A_\alpha (\hat\lambda^{-2}(0))_{\alpha\beta}A_\beta,\nn
\end{align}
where $\mathbf{j} = en_c\mathbf{v}_\mathrm{dr},$
$\mathbf{v}_\mathrm{dr}= m^{-1}\cdot\mathbf{Q},$ and
\begin{gather}
 n_c     = 2\int_{\eps_{\bp} < E_\mathrm{F}} \frac{d \bp}{(2\pi\hbar)^3}\nn\\
 m^{-1} = \frac{\D\int\frac{d\bp}{(2\pi\hbar)^3}m^{-1}_{\bp}
    \theta(E_{\mathrm{F}} -\eps_{\bp})
  }{\D
    \int\frac{d \bp}{(2\pi\hbar)^3}\theta(E_{\mathrm{F}} -\eps_{\bp})}
        =\frac{\D
    \int_{\eps_{\bp} = E_\mathrm{F}}
    \frac{dS_\bp}{(2\pi\hbar)^3v_\mathbf{p}}
    \mathbf{v}_{\bp} \otimes \mathbf{v}_{\bp}
       }{\D
    \int_{\eps_{\bp} < E_\mathrm{F}}
    \frac{d\bp}{(2\pi\hbar)^3}
       }.
\end{gather}
Whence the expressions for the penetration depth and the optical mass
tensor read
\begin{align}
\label{eq:lambda0}
(\hat\lambda^{-2}(0))_{\alpha\beta} & =
     \frac{e^2n_c}{\eps_0c^2} (m^{-1})_{\alpha\beta}
    =\frac{e^2 }{\eps_0c^2}2\nu_\mathrm{F}\la v_\alpha v_\beta\ra,\\
(m^{-1})_{\alpha\beta} &=
\frac{2\nu_{\mathrm{F}}}{n_c}\la v_\alpha v_\beta\ra,\\
\nu_{\mathrm{F}} & =
\D \int\frac{d \bp}{(2\pi\hbar)^3}\delta(E_{\mathrm{F}}
    -\eps_{\bp}).
\end{align}
The comparison with the Bernoulli effect considered in Sec.~3.1 tells
us that the effective mass of the Cooper pairs in the clean limit is
just twice the optical mass:
\begin{equation}
 M_\mathrm{CP}=2m,\quad n(T=0)=\frac{1}{2}n_c,
\end{equation}
and neglecting some subtleties, e.g., the appearance of hole pockets,
the total number of electrons corresponds to the total number of
charge carriers $n_c$ and $n_{\mathrm{tot}}.$

Formally, the effect of nonzero temperature reduces\cite{Kogan:02} to
introducing an additional multiplier $r_d$ in the averaging of the
velocity-velocity tensor at the Fermi surface, i.e.
\begin{equation}
(\hat\lambda^{-2}(T))_{\alpha\beta} =\frac{e^2
}{\varepsilon_0c^2}2\nu_\mathrm{F} \la v_\alpha v_\beta
r_d\left(\frac{\Delta_{\mathbf{p}}}{2\pi\kb T}\right)\ra,
\end{equation}
where
\begin{equation}
\label{eq:K-function}
r_d\left(\frac{\Delta_{\mathbf{p}}}{2\kb T}\right)
 = \sum_{n=0}^\infty \frac{2\pi\kb T\Delta_{\mathbf{p}}^2}
   {\left\{\Delta_{\mathbf{p}}^2+\varepsilon_n^2 \right\}^{3/2}}
\approx
\begin{cases}
 1, & \mbox{ for }\kb T \ll |\Delta_{\bp}|\\
 7\zeta(3)\left(\frac{\Delta_{\bp}}{2\pi\kb T}\right)^2,
    & \mbox{ for }|\Delta_{\bp}|\ll\kb T
\end{cases},\nn
\end{equation}
where $\varepsilon_n=(2n+1)\pi\kb T.$
The BCS multiplier $r_d$ can be easily derived using Matsubara
Green function method\cite{LL:9}, averaging over the Fermi surface
also can be easily derived using this technique. For
superconductors with strong coupling corrections we suggest an
interpolation formula which can be used before a detailed theory
to be developed. Having the solution of the Eliashberg equations
$Z_{n,\mathbf{p}}$, $\tilde\varepsilon_{n,\mathbf{p}}$ and $\tilde
\Delta_{n,\mathbf{p}}$ we can perform in \Eref{eq:K-function} the
substitution
$\varepsilon_n\rightarrow\tilde\varepsilon_{n,\mathbf{p}}=Z_{n,\mathbf{p}}\varepsilon_{n}$
and $\Delta_{n,\mathbf{p}}\rightarrow\tilde
\Delta_{n,\mathbf{p}}$. The procedure used by Kogan\cite{Kogan:02}
is related to further substitution in \Eref{eq:K-function}
$\tilde\varepsilon_{n,\mathbf{p}}\rightarrow\tilde\varepsilon_{n,\mathbf{p}}'
=\tilde\varepsilon_{n,\mathbf{p}}+1/\tau_\mathbf{p}$.
The function
\begin{equation}
r_d(z)  = z^2\sum_{n=0}^\infty
\left[z^2+\pi^2\left(n+1/2\right)^2\right]^{-3/2} \approx
\begin{cases}
     7\zeta(3)z^2/\pi^2, & z\ll1\\
     1,            & z\gg1
    \end{cases},
\end{equation}
gives implicitly the BCS expression\cite{bcs:57} for the current
response (for a simple derivation using temperature Green's functions
see, for example, \Rref{LL:9})

For multiband superconductors, like MgB$_2$, the general formula for
the penetration depth reads
\begin{align}
(\hat\lambda^{-2}(T))_{\alpha\beta} & =
    2 \nu_\mathrm{F}\frac{e^2 }{\varepsilon_0c^2}
    \sum_b c_b\, r_d\left(\frac{\Delta_{b}(T)}{2\pi\kb T}\right)
    \la v_\alpha v_\beta\ra_b,\nn\\
 & = \sum_b c_b (\hat\lambda_b^{-2}(0))_{\alpha\beta}\,
    r_d\left(\frac{\Delta_{b}(T)}{2\pi\kb T}\right),
\end{align}
where $\lambda_b^{-2}(0)$ is the contribution of the $b$th band at
$T=0.$ A careful fit of $\lambda(T)$ for MgB$_2$ could make it
possible to separate the influence of the $\pi$- and $\sigma$-bands.

Returning to the expression for the kinetic energy density,
\Eref{eq:OpticalResponse}, let us introduce the momentum of the pairs
$\mathbf{\Pi}=2\mathbf{Q},$ and the small-gap approximation
\[
  r_d(\mathbf{p})\approx 7\zeta(3)(|\Xi|\chi_{\bp}/2\pi\kb T)^2.
\]
Then for a clean superconductor one obtains
\begin{align}
\label{eq:Simplicio}
 w_{\mathrm{kin}} &=
    \frac{1}{2}\int_{\varepsilon_\mathbf{p}<E_\mathrm{F}}\frac{2d\mathbf{p}}{(2\pi\hbar)^3}
\mathbf{Q}\cdot
m^{-1}_\mathbf{p}\cdot\mathbf{Q}\,r_d(\mathbf{p})\\
&=\frac{1}{2}(\mathbf{\Pi}\Xi)^*\cdot\tilde
    M^{-1}\cdot(\mathbf{\Pi}\Xi)
=- a_0 [ \mathbf{D}\Xi(\mathbf{r})]^* \cdot \hat\xi^2 \cdot
    \mathbf{D}\Xi(\mathbf{r})\nn\\
&=-\frac{7\zeta(3)\hbar^2\nu_\mathrm{F}}{\left(4\pi\kb \Tc\right)^2}\,
 [\mathbf{D}\Xi(\mathbf{r})]^* \cdot
 \la \mathbf{v}\otimes \mathbf{v} \chi^2\ra \cdot
 \mathbf{D}\Xi(\mathbf{r}),\nn
\end{align}
where we have used the replacement
\begin{equation}
\mathbf{\Pi}\Xi \rightarrow -i\hbar \mathbf{D}\Xi(\mathbf{r})
 =\left(-i\hbar \frac{\partial}{\partial\mathbf{r}}
  -e^* \mathbf{A}(\mathbf{r})\right) \Xi(\mathbf{r}).
\end{equation}
Equation~(\ref{eq:Simplicio}) agrees with \Eref{eq:CoherenceLength}
for $x_c=0$ which is perhaps the simplest validation of the gradient
terms in the GL expansion.  In this way, for temperatures slightly
below \Tc\ we obtain
\begin{gather}
\tilde M_{\mathrm{clean}}^{-1}=\frac{7\zeta(3)\nu_{\mathrm{F}}}
{8\pi^2(\kb \Tc)^2}\la \mathbf{v}\otimes \mathbf{v} \chi^2\ra,\\
\hat\lambda_{\mathrm{clean}}^{-2}(T)\approx\frac{4e^2\nu_{\mathrm{F}}}
{\varepsilon_0c^2}
\frac{\la\chi^2\ra \la \mathbf{v}\otimes \mathbf{v} \chi^2\ra}{\la\chi^4\ra}
\frac{\Tc-T}{\Tc}.
\end{gather}
For isotropic-gap superconductors, $\chi=\mathrm{const},$ and
arbitrary band anisotropy this equation together with
\Eref{eq:lambda0} gives
\begin{equation}
\hat\lambda_{\mathrm{clean}}^{-2}(0)=
\left.
-\frac{1}{2}\frac{d}{dT}
\hat\lambda_{\mathrm{clean}}^{-2}(T)\right|_{\Tc}.
\label{azmcy}
\end{equation}

\pagebreak[3]

\section{Conclusions}

We demonstrated that temperature dependence of the penetration
depth\cite{Kogan:02} in clean superconductors can be presented and
an inserting of an additional BCS factor $r_d$ in the formula for
far infrared skin depth of the metal. We conclude that London-Hall
coefficient is temperature- and disorder independent, and the
Bernoulli effect is an adequate method for determination of ${\cal
R}_{\mathrm{H}}.$ Given the penetration depth and the London-Hall
coefficient at $T=0$ we can determine the effective mass of the
superfluid particles. Measuring the Bernoulli effect for thin and
thick films, see Eq.~(\ref{eq:bulk}) and Eq.~(\ref{eq:BTH}) we
conclude that London penetration depth $\lambda(T)$ also can
determined by purely electrostatic measurements. This hydrodynamic
mass $M_\mathrm{CP}$ which parameterizes the relation between
${\cal R}_{\mathrm{H}}$ and $\lambda(0)$ is in the clean limit the
optical mass extrapolated to zero frequency. But for disordered
superconductors, irradiated thin films, for example, investigation
of the disorder dependence of the Bernoulli signal can give an
important information for the disorder renormalization of the
effective Cooper pair mass. On Fig.~\ref{fig:1} is presented the
result for an isotropic gap superconductor but for a different gap
anisotropies the curves will be different. In such a way we are
coming to the conclusion that the impurity and temperature
dependence of the Bernoulli signal is an important property which
can be used to check the comparison between the theory and the
experiment. For high-\Tc\ superconductors the Bernoulli effect
gives the possibility to evaluate the vortex
charge.\cite{Mishonov:00} We emphasize that Bernoulli effect and
other electric field effects in superconductors, e.g., the surface
Hall effect,\cite{Mishonov:99} gives a complete set of parameters
describing the low-frequency current response of superconductors.
We conclude also that bulk condensation energy of a superconductor
can also be determined by electrostatic polarization created by a
sudden heating due to Eq.~(\ref{Eq:CPD T}).

\nonumsection{Acknowledgments}
\noindent

This work was partially supported by the Flemish Government Programme
GOA.

%
\nonumsection{References}


\begin{thebibliography}{99}
%
\bibitem{LLV} L.~D.~Landau and E.~M.~Lifshitz, \textit{Course on
Theoretical Physics, Vol. 5, Statistical Physics, Part~1} (Pergamon
Press, Oxford, 1980), Ch.~XIV.
%
\bibitem{GL} V.~L.~Ginzburg and L.~D.~Landau,
\jour{Zh.~Exp.~Teor.~Fiz.}{20}{1064}{1950}.

\bibitem{LL:9} E.~M.~Lifshitz and L.~Pitaevskii, \textit{Course on
Theoretical Physics, Vol. 9, Statistical Physics, Part~2} (Pergamon
Press, Oxford, 1981), Sec.~45, Sec.~51.

\bibitem{bcs:57} J.~Bardeen, L.~N.~Cooper, and J.~R.~Schrieffer,
\jour{\PR}{108}{1175}{1957}, Eq.~(3.46).

\bibitem{Pokrovsky:96} S.~V.~Pokrovsky and V.~L.~Pokrovsky,
\jour{\PR}{B54}{13275}{1996};
T.~M.~Mishonov, E.~S.~Penev, J.~O.~Indekeu, V.~L.~Pokrovsky
cond-mat/0209342, ``Specific Heat Discontinuity in Impure Two-Band
Superconductors''.

\bibitem{Kogan:02} V.G.~Kogan, \textit{Macroscopic anisotropy in
superconductors with anisotropic gaps}, cond-mat/0204038, Eq.~(13),
Phys. Rev. B \textbf{66}, 020509 (2002);
\textit{Free energy and torque for superconductors with different
anisotropy of $H_{c2}$ and $\lambda$}, cond-mat/0208362,
Phys. Rev. Lett. \textbf{89}, 237005 (2002).

\bibitem{Pippard:53} A.~B.~Pippard,
\jour{Proc. Roy. Soc. London}{A216}{547}{1953}; T.~E.~Faber and
A.~B.~Pippard, \jour{Proc. Roy. Soc. London}{A231}{336}{1955}.


\bibitem{Ketterson:99} J.~B.~Ketterson and S.~N.~Song,
\textit{Superconductivity} (Cambridge University Press, Cambridge,
1999) Sec.~45.4, p.~330, Eq.~(45.19) and Sec.~33, p.~10, Eq.~(3.14).

\bibitem{Tinkham:96} M.~Tinkham, \textit{Introduction to
Superconductivity}, 2nd edition (McGraw-Hill, New York, 1996),
Sec.~1.7, p.~7, Sec.~3.10.4, p.~97, Eq.~(3.123b), Sec.~4.1, p.~115,
and Sec.~4.4, p.~124.

\bibitem{Mishonov:94a} T.~M.~Mishonov,
\jour{\PR}{B50}{4004}{1994}.
%
\bibitem{Mishonov:94b} T.~M.~Mishonov,
\jour{\PR}{B50}{4009}{1994}, Fig.2.
%
\bibitem{Mishonov:91} T.~M.~Mishonov.
\jour{\PRL}{67}{3195}{1991}.
%
\bibitem{Fiory:91} A.~T.~Fiory, A.~F.~Hebard, R.~E.~Eick,
P.~M.~Mankiewich, R.~E.~Howard, and M.~L.~O.~Malley.
\jour{\PRL}{67}{3196}{1991}.
%
\bibitem{Mishonov:99} T.~Mishonov and N.~Zahariev, \jour{Superlattices
and Microstructures}{26}{57}{1999}.
%
\bibitem{Mishonov:00} T.~M.~Mishonov, ``In search for the vortex charge
and the Cooper pair mass'', in \textit{Superconducting and Related
Oxides: Physics and Nanoengineering} IV, D.~Pavuna and I.~Bozovic,
Editors, Proceedings of \textit{SPIE} Vol. \textbf{4058}, pp. 97--108
(2000), Orlando, Florida, USA, 24--28 April 2000; cond-mat/0004286 (19
April 2000); T.~Mishonov, D.~Nikolova, K.~Kaloyanov, Sl.~Klenov,
``Bernoulli effect in the quantum Hall regime and type-II
superconductors in magnetic field'' in \textit{High-\Tc\
Superconductors and Related Materials: Material Science, Fundamental
Properties, and Some Future Electronic Applications}, edited by
S.-L.~Drechsler and T.~Mishonov (Kluwer, Dordrecht, 2001).

\bibitem{Kumagai:01} K.~I.~Kumagai, K.~Nozaki and Y.~Matsuda,
\jour{\PR}{B63}{144502}{2001}.

\end{thebibliography}
\end{document}